\documentclass[twoside,twocolumn,10pt]{article}
\usepackage{amsmath}
\usepackage{balance}
\usepackage[format=plain,justification=justified,singlelinecheck=false,font={stretch=1.125,small},labelfont=bf,labelsep=space]{caption}
\usepackage{charter}
\usepackage{fnpos}
\usepackage[left=1.5cm, right=1.5cm, top=1.8cm, bottom=2.0cm]{geometry}
\usepackage{graphicx} 
\usepackage{hhline}
\usepackage{hyperref}
\usepackage[none]{hyphenat}
\usepackage{lastpage}
\usepackage[sort&compress,numbers]{natbib} 
\usepackage{sectsty}
\usepackage{setspace}
\usepackage{siunitx}
\usepackage{subcaption}
\usepackage{tabularx}
\usepackage[compact]{titlesec}
\usepackage[sort]{cleveref}

\DeclareSIUnit\rpm{rpm}
\crefname{table}{Table}{Tables}
\crefname{equation}{Eq.}{Eqs.}
\Crefname{figure}{Fig.}{Figs.}
\setcitestyle{square}

\makeatletter
\renewcommand{\@seccntformat}[1]{%
	\ifcsname prefix@#1\endcsname
	\csname prefix@#1\endcsname
	\else
	\csname the#1\endcsname\quad
	\fi}
\newcommand\prefix@section{}
\newcommand\prefix@subsection{}
\makeatother

\makeatletter
\DeclareRobustCommand*\textsubscript[1]{%
  \@textsubscript{\selectfont#1}}
\def\@textsubscript#1{%
  {\m@th\ensuremath{_{\mbox{\fontsize\sf@size\z@#1}}}}}
\makeatother

\makeatletter
\renewcommand\LARGE{\@setfontsize\LARGE{15pt}{17}}
\renewcommand\Large{\@setfontsize\Large{12pt}{14}}
\renewcommand\large{\@setfontsize\large{10pt}{12}}
\renewcommand\footnotesize{\@setfontsize\footnotesize{7pt}{10}}
\makeatother

\sectionfont{\rmfamily\Large}
\titlespacing*{\section}{0pt}{10pt}{4pt}

\begin{document}
\twocolumn[
  \begin{@twocolumnfalse}
\vspace{1in}

\noindent\LARGE{\textbf{Continuously Phase-Modulated Standing Surface Acoustic Waves for Separation of Particles and Cells in Microfluidic Channels Containing Multiple Pressure Nodes}}

\strut

\noindent\large{Junseok Lee,\textit{$^{a\dag}$} Chanryeol Rhyou,\textit{$^{a\dag}$} Byungjun Kang,\textit{$^{a}$} and Hyungsuk Lee$^\ast$\textit{$^{a}$}}

\strut

\noindent\normalsize{\textbf{Abstract} \newline This paper describes continuously phase-modulated standing surface acoustic waves (CPM-SSAW) and its application for particle separation in multiple pressure nodes. A linear change of phase in CPM-SSAW applies a force to particles whose magnitude depends on their size and contrast factors. During continuous phase modulation, we demonstrate that particles with the target dimension are translated in the direction of moving pressure nodes, whereas smaller particles show oscillatory movements. The rate of phase modulation is optimized for separation of target particles from the relationship between mean particle velocity and period of oscillation. The developed technique is applied to separate particles of the target dimension from the particle mixture. Furthermore, we also demonstrate human keratinocyte cells can be separated in the cell and bead mixture. The separation technique is incorporated with a microfluidic channel spanning multiple pressure nodes, which is advantageous over separation in a single pressure node in terms of high throughput.}
 \end{@twocolumnfalse} \vspace{\baselineskip}
  ]

\renewcommand*\rmdefault{bch}\normalfont\upshape
\rmfamily

\footnotetext{\textit{$^{a}$~School of Mechanical Engineering, Yonsei University, Seoul, Korea}}
\footnotetext{$^\dag$~These authors contributed equally to this work.}
\footnotetext{$^\ast$~Author to whom correspondence should be addressed. Corresponding email: hyungsuk@yonsei.ac.kr}

The separation of targeted objects is an essential first step in analyzing biological and chemical samples \cite{yung09,li15,devendran16}.
In addition to passive separation techniques such as inertial microfluidics, active methods using electromagnetics, optics, or acoustics have been developed to improve the separation  efficiency \cite{sajeesh14,lenshof10}.
Recently, particle manipulation using surface acoustic waves (SAW) is emerging because of its high energy efficiency, label-free characteristics, and applicability to sound-absorbing materials such as polydimethylsiloxane (PDMS) \cite{ding13}.
Two major types of SAW are travelling SAW (TSAW) and standing SAW (SSAW), both of which are shown to have the ability to separate particles \cite{destgeer13,destgeer2015microchannel,skowronek2015surface,lee2015acoustic,ren2015high,nawaz2015acoustofluidic,nam2011separation,jo2012active}.

\begin{figure*}[!t]
	\centering%
	\includegraphics[width=2\columnwidth]{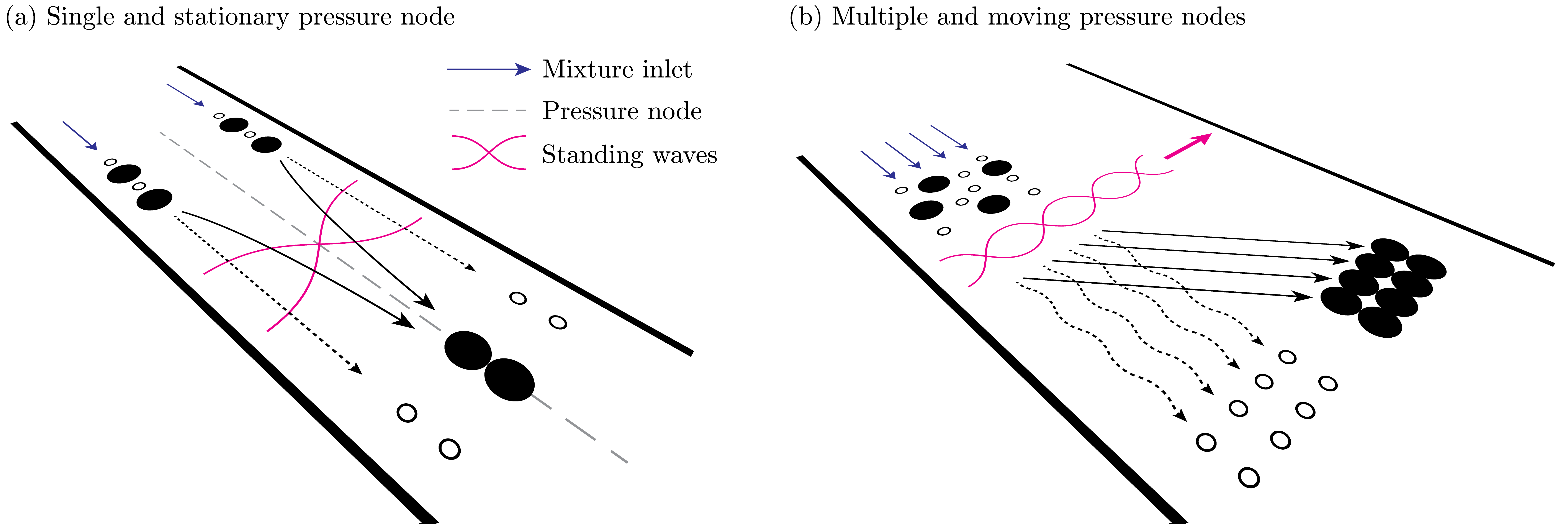}
	\begin{subfigure}{0\textwidth}%
		\phantomcaption%
		\label{fig:convSSAW}%
	\end{subfigure}%
	\begin{subfigure}{0\textwidth}%
		\phantomcaption%
		\label{fig:MPNsucceeds}%
	\end{subfigure}%
	\caption{
		\textbf{Comparison of SSAW-based particle separation techniques.}
		\subref{fig:convSSAW} In separation utilizing SSAW, the acoustic radiation force drives the larger particles to pressure nodes faster than smaller ones. 
		When the location of pressure node is fixed, particles can be separated by the difference in arrival time depending on the size of particle.
		The channel width should be limited by half-wavelength.
		\subref{fig:MPNsucceeds} In the proposed technique, pressure nodes are translated by phase modulation. Particles with the target size are continuously displaced in the direction of the moving pressure nodes, while non-target particles travel along the flow direction with oscillatory motion. 
		Widening channels to contain multiple pressure nodes is advantageous for massive particle separation.
	}    
\end{figure*}

The separation of particles using SSAW is achieved by an acoustic radiation force that moves particles toward the nearest pressure nodes or antinodes. As the magnitude of force depends on the volume of particles, particles of different sizes initially placed at the same positions are attracted to pressure nodes or antinodes at different speeds, and can be separated by the difference in arrival time [\Cref{fig:convSSAW}]. However, once particles are aligned at their pressure nodes or antinodes, acoustic radiation force on particles become zero preventing their transverse displacement. Therefore, the separation technique is only applicable to a narrow channel. The channel width is limited to contain a single pressure node, hindering massive separation with high throughput.

In this paper, we describe a continuously phase-modulated SSAW (CPM-SSAW) that can be applied to separate particles in a channel spanning more than one pressure node for high throughput [\Cref{fig:MPNsucceeds}].
Pressure nodes are continuously displaced by modulating phase at the rate to exert the maximized acoustic radiation force on particles with a target size. 
Due to the balance between drag and acoustic radiation forces, they are displaced at the constant velocity equal to that of the moving pressure nodes.
In response to the phase modulation adjusted for the target particle, non-target particles show oscillatory movements with a mean velocity smaller than the speed of moving pressure nodes.
A theoretical analysis shows that the mean velocity and the oscillation period of particles in CPM-SSAW are determined by their size and contrast factor. We obtain the optimal rate of phase change for separation through the nondimensional analysis. By modulating the phase at the optimal rate, we can separate particles and cells of the target dimension from a mixture. This technique is particularly advantageous in increasing separation efficiency, because the channel width is not limited to half-wavelength of SSAW.

Two dominant forces in acoustofluidics are acoustic radiation induced by the scattering of waves, and drag. The acoustic radiation force of standing waves $\mathbf{F}_r$ is given by
\begin{subequations}
	\begin{gather}
	\mathbf{F}_r  =  F_0 \sin \left( {2kx} \right) \mathbf{ \hat e}_x , \label{eqn:ARF} \\
	F_0  = \frac{{\pi p_0^2{V_c}{\beta _w}}}{{2\lambda }}\varphi,\qquad
	\varphi  = \frac{{5{\rho _c} - 2{\rho _w}}}{{2{\rho _c} + {\rho _w}}} - \frac{{{\beta _c}}}{{{\beta _w}}} \label{eqn:ARF0}
	\end{gather}
\end{subequations}
where $k$, $\lambda$, $p_0$, $V_c$, $\rho_c$, $\rho_w$, $\beta_c$, $\beta_w$, and $\varphi$ are the wavenumber, wavelength, acoustic pressure, particle volume, particle density, fluid density, particle compressibility, fluid compressibility, and a contrast factor \cite{bruus12}.
According to \cref{eqn:ARF}, particles with positive and negative contrast factors are attracted to pressure nodes and antinodes, respectively.
As the first term of the contrast factor is generally positive due to the large multiplier in front of $\rho_c$, for most cases, its sign depends on the ratio of compressibility. For examples, rigid particles have positive contrast factors, and soft cells have negative ones.
The drag force $\mathbf{F}_d$ is determined by Stokes' drag, which is given by
\begin{subequations}	
	\begin{align}
	\mathbf{F}_d & =  - b\left(\mathbf{v-u}\right), \label{eqn:stokes}\\
	b & = 6\pi a\eta, \label{eqn:stokes0}
	\end{align}
\end{subequations}
where $\mathbf{v}$, $\mathbf{u}$, $a$, and $\eta$ are the particle velocity, flow velocity, particle radius, and fluid viscosity \cite{bruus11}.
We assume the motion is one-dimensional in a quiescent fluid, ${\mathbf{u}}={\mathbf{0}}$. As the phase modulation of an SSAW to $\phi$ displaces pressure nodes by $\phi/2k$, the acoustic radiation force of the phase-modulated SSAW can be estimated by substituting $x \to x - \phi /2k$ in \cref{eqn:ARF} \cite{meng11}. Hence, $F_r = F_0 \sin \left( {2kx - \phi } \right).$
If the rate of phase modulation is sufficiently small compared to the frequency of waves, the particle motion can be described by
\begin{equation} \label{eqn:motion}
b\frac{{dx}}{{dt}} = F_0 \sin \left( {2kx - \phi } \right),
\end{equation}
as the inertia of a particle can be neglected in microfluidics \cite{nama15}.

\begin{figure}[!t]
	\centering%
	\includegraphics[width=\columnwidth]{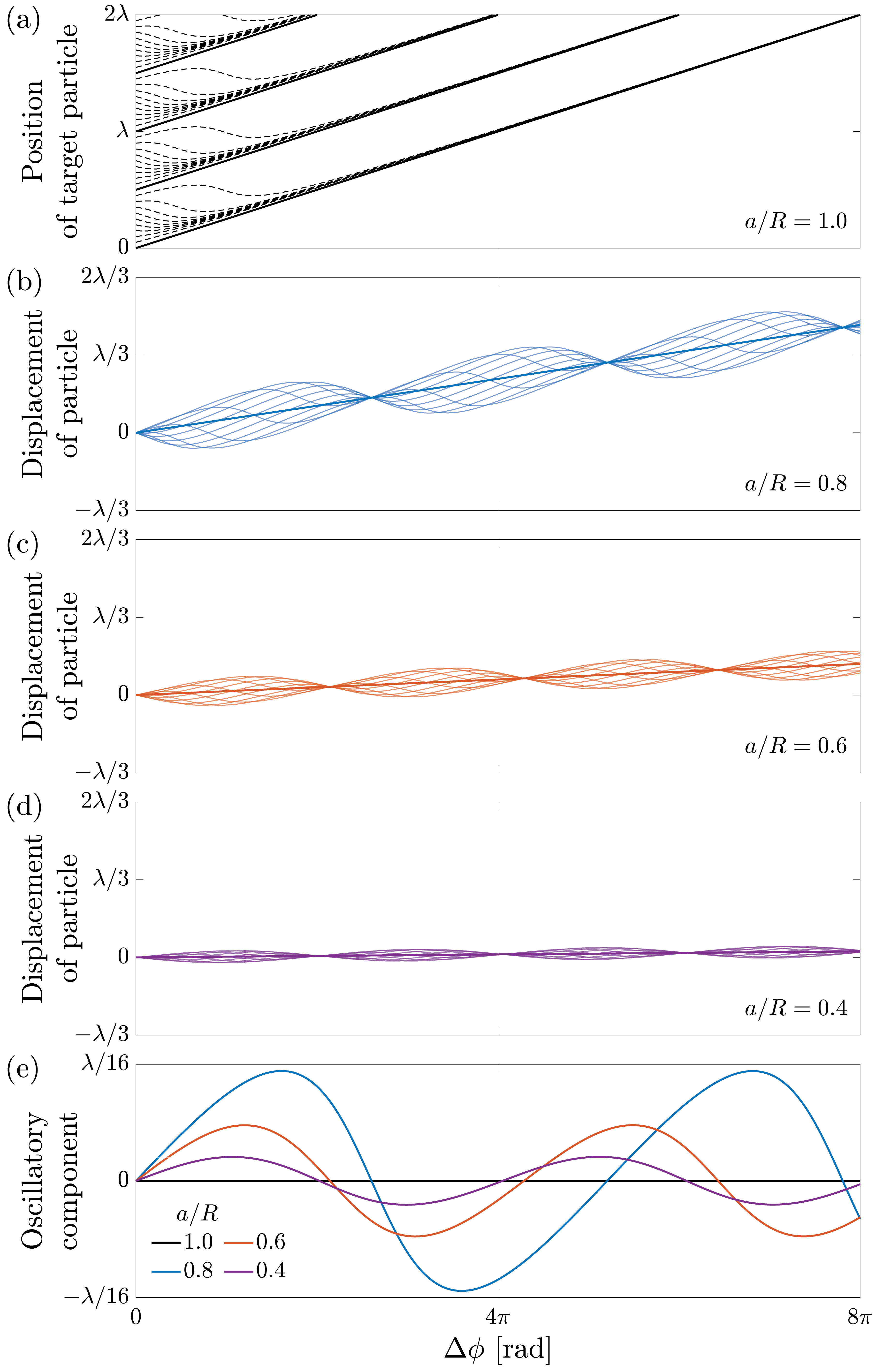}%
	\begin{subfigure}{0\textwidth}%
		\phantomcaption%
		\label{fig:fig1_initial}%
	\end{subfigure}%
	\begin{subfigure}{0\textwidth}%
		\phantomcaption%
		\label{fig:fig1_dia12}%
	\end{subfigure}%
	\begin{subfigure}{0\textwidth}%
		\phantomcaption%
		\label{fig:fig1_dia09}%
	\end{subfigure}%
	\begin{subfigure}{0\textwidth}%
		\phantomcaption%
		\label{fig:fig1_dia06}%
	\end{subfigure}%
	\begin{subfigure}{0\textwidth}%
		\phantomcaption%
		\label{fig:fig1_period}%
	\end{subfigure}%
	\caption{%
		\textbf{Responses of particles to CPM-SSAW depending on its size.}
		\subref{fig:fig1_initial} Position of target particles with radius $R$ at various initial positions during a linear phase modulation as in \cref{eqn:phase}. Traces converge to the thick solid lines, implying that mean velocities of particles are independent of their initial positions.
		\subref{fig:fig1_dia12}, \subref{fig:fig1_dia09}, \subref{fig:fig1_dia06} Displacement of non-target particles of various radii ($0.8R$, $0.6R$, and $0.4R$) with various initial positions, respectively. The mean velocities of each size particle are not influenced by initial positions. Compared to target particles with radius $R$ shown in \Cref{fig:fig1_initial},  displacements of smaller particles delay and exhibit oscillatory movement about center lines (thick lines).
		\subref{fig:fig1_period} The oscillating components ($\chi$) decomposed from the positions of particles differ depending on the particle size. Both period and amplitude decrease as the particle size decreases.
	}
\end{figure}            

Suppose that the target particle is of radius $R$, and the phase is adequately modulated so that the maximum acoustic radiation force is continuously applied to the particle.
Then, \cref{eqn:motion} separates into
\begin{equation} \label{eqn:systemsofeqn}
b\frac{{dx}}{{dt}} = F_0,  \qquad
\sin \left( {2kx - \phi } \right) = 1.
\end{equation}
Therefore, the position of the particle and the phase to be modulated are given by
\begin{equation}
x  = \overline{V}t+x_0,\qquad
\phi  = \frac{2 k F_0}{b}t +2 k x_0 -\frac{\pi}{2}, \label{eqn:phase}
\end{equation}
where $x_0$ is the initial position of the particle and $\overline{V}$ represents $F_0/b$. Note that the phase to be modulated is a linear function with respect to time; thus, the pressure nodes translate at a uniform speed that is equal to the speed of the target particle.

{\renewcommand{\arraystretch}{1.125}
	\begin{table}[t]
		\centering
		\begin{tabularx}{\columnwidth}{Xcrl}
			\hline & & & \vspace{-1em}\\
			\textbf{Material Properties}  & & & \\ \hhline{====}
			Water  & & & \\
			\hline
			Viscosity & $\eta$ & 0.98 & \si{\milli\pascal\second} \\
			Compressibility & $\beta_w$ & 448 & \si{\per\tera\pascal} \\
			Density & $\rho_w$ & 998 & \si{\kilogram\per\cubic\meter} \\
			\hline
			Polystyrene Particles & & & \\
			\hline
			Compressibility & $\beta_c$ & 249 & \si{\per\tera\pascal} \\
			Density & $\rho_c$ & 1050 & \si{\kilogram\per\cubic\meter} \\
			Diameter of the Target Particle & $2R$ & 15 & \si{\micro\meter} \\
			\hline
			Acoustic Waves & & & \\
			\hline
			Acoustic Pressure & $p_0$ & 100 & \si{\kilo\pascal} \\
			Wavelength & $\lambda$ & 280 & \si{\micro\meter} \\
			\hline
			Channel Geometry & & & \\
			\hline
			Width & $w$ & 5 & \si{\milli\meter} \\
			Height & $h$ & 5, 0.5 &  \si{\milli\meter} \\			
			Length & $L$ & 12 & \si{\milli\meter} \\
			\hline
			Poiseuille Flow & & & \\
			\hline
			Average Velocity & $\overline{U}$ & 400 & \si{\micro\meter\per\second} \\
			\hline
		\end{tabularx}
		\captionsetup{width=\columnwidth}
		\caption{Numerical values used for the Runge-Kutta method and COMSOL Multiphysics\textsuperscript{\textregistered}.\label{table:table1}}
	\end{table}}

We investigate how differently particles initially located at arbitrary positions different from $x_0$  respond to the phase change given by \cref{eqn:phase} for $x_0$. In this case, the right-hand side of \cref{eqn:motion} no longer yields a constant term as in \cref{eqn:systemsofeqn}, and the equation of motion can be solved using numerical analysis with material properties provided in \cref{table:table1}. When the phase increases linearly, positions of the particles converge to multiple lines that are equally spaced by a half wavelength due to the periodicity of the acoustic radiation force [\Cref{fig:fig1_initial}]. Note that the initial positions do not change the mean velocity, which simplifies the subsequent theoretical development. This result suggests that the CPM-SSAW technique is capable of separating particles that are randomly distributed in an extensive area, regardless of their initial positions. In particular, the independence of the initial positions endows a degree of robustness against disturbances such as flow deflection and friction between particles and the channel wall. Even if a particle fails to be displaced by the moving pressure node, it can be displaced by the following one.

We also analyze the response of non-target particles of radius $a$ to the continuous phase modulation. Similar to \cref{eqn:motion}, the equation of motion for non-target particles is given by
\begin{equation} \label{eqn:motionb}
b'\frac{{dx'}}{{dt}} = F'_0 \sin ( {2kx' - \phi } ),
\end{equation}
where primes denote quantities related to non-target particles.
Note that $\phi$ is the estimated phase for the target particle from \cref{eqn:phase}.
We find that particles with a radius smaller than the target particle exhibit periodic oscillation with a smaller mean velocity compared to the speed of moving pressure nodes. The mean velocities are independent of initial positions, as is the case with target particles [Figs. \ref{fig:fig1_dia12}, \subref{fig:fig1_dia09}, and \subref{fig:fig1_dia06}].

\begin{figure}[!t]
	\centering
	\includegraphics[width=\columnwidth]{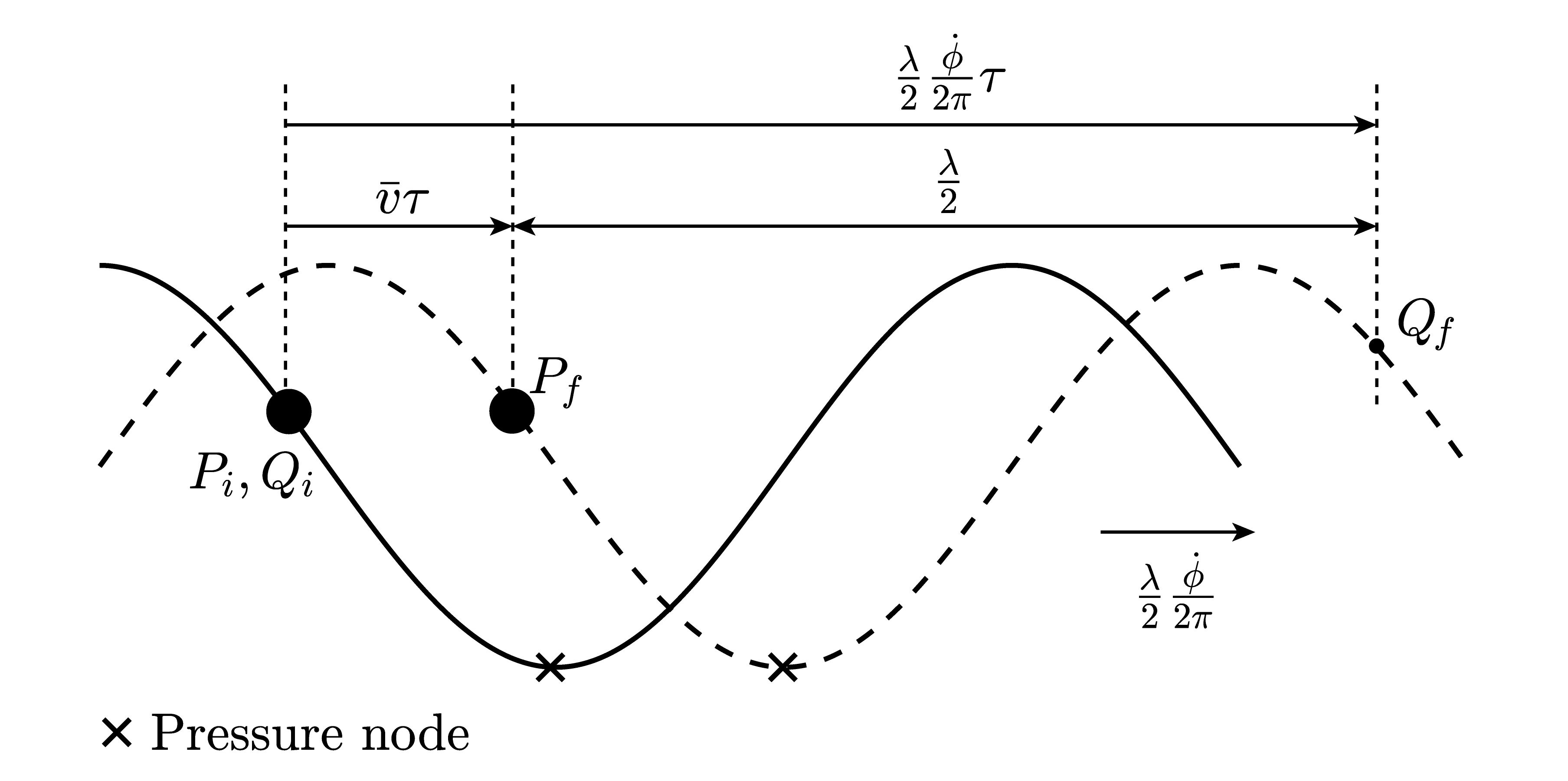}
	\caption{
		Moving Gor'kov potentials at $t=t_0$ (solid line) and $t=t_0+\tau$ (dashed line). $P_i$ and $Q_i$ are the position of the particle and Gor'kov potential at $t=t_0$, respectively. While Gor'kov potential travels from $Q_i$ to $Q_f$ during a period, the particle only travels from $P_i$ to $P_f$.		
		\label{sFig:sFig_illustration}
	}
\end{figure}

The oscillating component can be further analyzed by decomposing the displacement as $x' = \overline v t + \chi$, where $\overline v$, $\chi$ are the mean velocity and the oscillating component, respectively. The oscillating component is obtained by subtracting $\overline{v} t$ from $x'$ [\Cref{fig:fig1_period}]. Both the amplitude and period of oscillation decrease as the particle size decreases, \emph{i.e.}, as the mean velocity decreases. The relationship between the mean velocity $\overline v$ and the period $\tau$ can be derived from the definitions of period and mean velocity. During the period $\tau$, a particle with a mean velocity $\overline v$  travels $\overline v \tau$ from $P_i$ to $P_f$, while the Gor'kov potential translates $\dot \phi / 2\pi \times \lambda /2 \times \tau$ from $Q_i$ to $Q_f$ [\Cref{sFig:sFig_illustration}]. According to the definition of the period, the difference in positions $P_f$ and $Q_f$  should be $\lambda /2$. Thus, $\overline v$ and $\tau$ are related by
\begin{equation}  \label{eqn:period}
\left( {\frac{{\dot \phi }}{{2\pi }}\frac{\lambda }{2} - \overline v} \right)\tau = \frac{\lambda }{2}.
\end{equation}

As the initial position does not contribute to the mean velocity, we assume the target particle is initially placed at $x_0$ such that $\phi(0)=0$, which simplifies \cref{eqn:phase} to
\begin{equation} \label{eqn:phase0}
\phi  = \frac{2 k F_0}{b}t = \dot \phi t,
\end{equation}
where $\dot \phi = 2kF_0/b$.
Using \cref{eqn:phase0}, \cref{eqn:motionb} can be written in dimensionless form as
\begin{subequations}
	\begin{gather}
	\frac{{dX^\ast}}{{dT^\ast}} = \sin \left( {\rho^\ast X^\ast - T^\ast} \right) , \label{eqn:ndMotion} \\
	X^\ast  = \frac{{b'\dot \phi }}{{{F'_0}}}x', \qquad
	T^\ast  = \dot \phi t, \qquad
	\rho^\ast  = \frac{{2k{F'_0}}}{{b'\dot \phi }}, \label{eqn:ndVariables}
	\end{gather}
\end{subequations}
where $\ast$ denotes a dimensionless quantity.
Hereby, $X^\ast$ and $T^\ast$ are the dimensionless position and time of a particle, respectively, and $\rho^\ast$ is another dimensionless parameter.

From \cref{eqn:ndMotion}, we shall show that the sign of contrast factor does not change the mean velocity, and accordingly, the period.
Suppose two identical particles with the contrast factors of opposite signs $\varphi'$  and $-\varphi'$. Assuming that $\varphi'$ is positive, we denote variables relevant to a negative contrast factor, $-\varphi'$, by a subscript minus sign ($-$). Then, ${F'_{0-}} = -{F'_0}$ and $\rho_-^\ast = -\rho^\ast$. From \cref{eqn:ndMotion}, the nondimensionalized equation for $\rho_-^\ast$ becomes 
\begin{subequations}
	\begin{align}
	\frac{{dX_-^\ast}}{{dT_-^\ast}} & = \sin \left( {\rho_-^\ast X_-^\ast - T_-^\ast} \right), \\
	\frac{{d(-X_-^\ast)}}{{d(T_-^\ast-\pi)}} & = \sin \left( {(-\rho_-^\ast)(-X_-^\ast) - (T_-^\ast-\pi)} \right). \label{sEqn:trigonometric}
	\end{align}
\end{subequations}
The trigonometric identity $-\sin(x)=\sin(x+\pi)$ is used in \cref{sEqn:trigonometric}. By comparing \cref{sEqn:trigonometric} with \cref{eqn:ndMotion},
\begin{equation} \label{sEqn:TXRelation}
T_-^\ast = T^\ast + \pi,\qquad X_-^\ast = -X^\ast. 
\end{equation}
It should be noted that \cref{sEqn:TXRelation} are true only if the initial conditions for $X^\ast$ and $X_-^\ast$ are properly related. For example, the initial condition for the equation for $\rho_-^\ast$ that $X_-^\ast=0$ at $T_-^\ast=0$ corresponds to the initial condition for the equation for $\rho^\ast$ that $X^\ast=0$ at $T^\ast=-\pi$. In other words, $X^\ast=X_0^\ast$ at $T^\ast=0$ for a certain $X_0^\ast$. Finally, by substituting \cref{eqn:ndVariables} into \cref{sEqn:TXRelation}, we obtain
\begin{subequations}
	\begin{align} 
	\frac{b_-'x_-'}{F_{0-}'}  & = -\frac{b'x'}{F_0'}, \\
	x_-' & = x'. \label{sEqn:xequalxprime}
	\end{align}
\end{subequations}
Again, note that \cref{sEqn:xequalxprime} is true only if the initial condition for $X^\ast$ is properly given as $X_0^\ast$. It means that the trajectories of particles with negative contrast factors are identical to those of particles with positive ones that are initially placed at some certain locations. Because an initial position does not affect the mean velocity, particles with negative contrast factors exhibit the same mean velocity as particles with positive ones.

\begin{figure}[!t]
	\centering
	\includegraphics[width=\columnwidth]{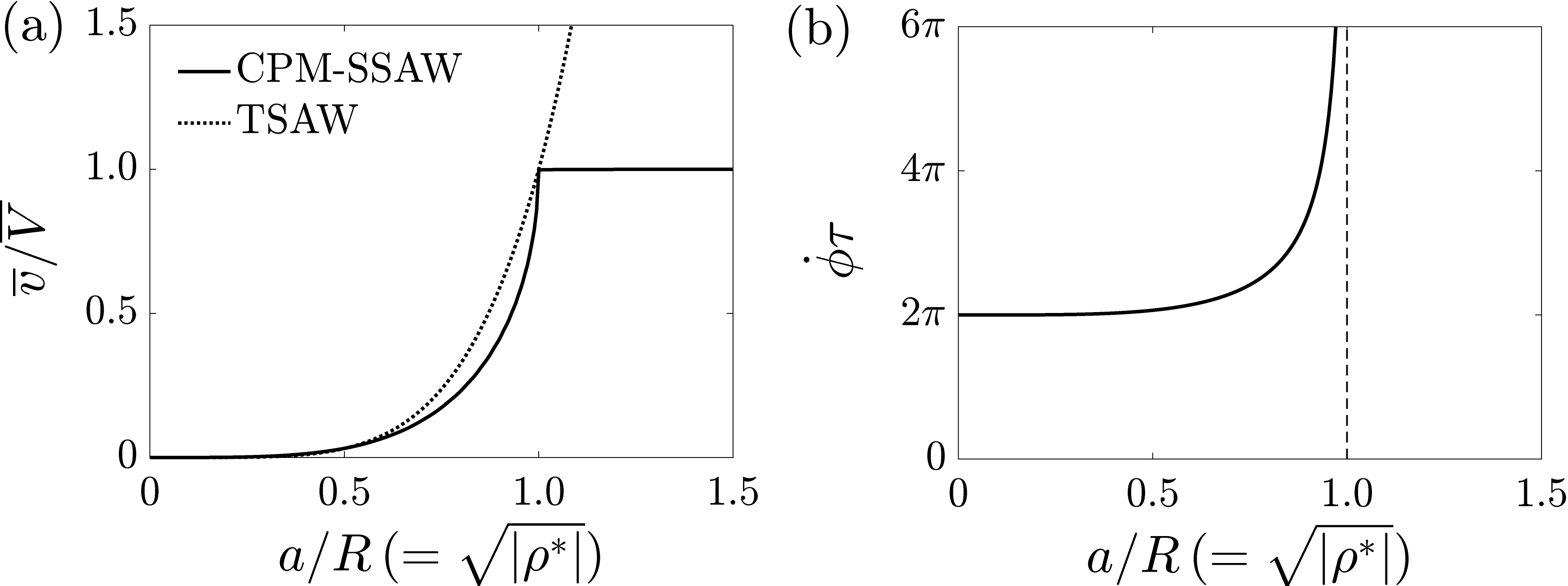}%
	\begin{subfigure}{0\textwidth}%
		\phantomcaption%
		\label{fig:fig1_normalized}%
	\end{subfigure}%
	\begin{subfigure}{0\textwidth}%
		\phantomcaption%
		\label{fig:fig1_normalizedPeriod}%
	\end{subfigure}%
	\caption{
		\subref{fig:fig1_normalized} The ratio of mean velocity of non-target particle to target particle ($\overline v/\,\overline V\,$) as a function of the radius ratio ($a/R$). The slope of curve at $a/R=1$ for CPM-SSAW is steeper than that for TSAW, which is advantageous for particle separation with high selectivity.
		\subref{fig:fig1_normalizedPeriod} Scaled period of oscillation as a function of particle size relative to $R$. For the target particle and larger ones ($a/R\geq 1$), the period diverges to infinity indicative of no oscillation in displacement. As the size of particle decreases, the oscillation period decreases significantly.
	}
\end{figure}

Ratios of radius can be estimated, using \cref{eqn:ARF0,eqn:stokes0,eqn:phase0}, as
\begin{align}
	{{\rho ^\ast}} & = {\frac{{\frac{{2k{F_0}^\prime }}{{b'}}}}{\dot \phi}} = {\frac{{\frac{{2k{F_0}^\prime }}{{b'}}}}{{\frac{{2k{F_0}}}{b}}}} = \frac{a^2}{R^2} {\frac{\varphi'}{\varphi}} \nonumber \\
	& = {\rm{sgn}}\left( {\frac{{\varphi '}}{\varphi }} \right)\left( {\frac{{a\sqrt {|\varphi '/\varphi |} }}{R}} \right)^2.
	\label{sEqn:detailedRhoStar}
\end{align}
As we only discuss the mean velocity and the period, which are independent of the sign of a contrast factor, one can neglect the signs of contrast factors to obtain
\begin{equation} \label{eqn:ndRadius}
\frac{a_e}{R}=\sqrt{|\rho^\ast|},
\end{equation}
where $a_e$ is an effective radius defined as $a\sqrt {\left| {\varphi '/\varphi } \right|}$.
Similarly, the ratio of velocity becomes
\begin{equation}
{\rho ^\ast}\frac{{\Delta {X^\ast}}}{{\Delta {T^\ast}}} = \frac{{\frac{{2k{F_0}^\prime }}{{b'}}}}{{\frac{{2k{F_0}}}{b}}}\frac{{\Delta \left( {\frac{{b'\dot \phi }}{{{F_0}^\prime }}x'} \right)}}{{\Delta \left( {\dot \phi t} \right)}} = \frac{1}{{\frac{{{F_0}}}{b}}}\frac{{\Delta x'}}{{\Delta t}} = \frac{\overline v}{\,\overline V\,}, \label{sEqn:velocityRatio}
\end{equation}
where $\overline{V}$ and $\overline{v}$ denote the mean velocity of the target and non-target particle, respectively.
A plot for the mean velocity with respect to the radius ratio reveals that smaller particles are displaced far more slowly than the target particles [\Cref{fig:fig1_normalized}]. When the radius ratio is 0.4, the mean velocity is only 1.3\% of that of target particle. For particles larger than the target particle, the displacement velocity is limited by the speed of the moving pressure node and it is equal to that of target particle. CPM-SSAW is similar to a TSAW in which an applied force to a particle depends on radius to the sixth power \cite{tan07}. However, it can be differentiated by the existence of the velocity limit in particle displacement.

Separation of particle can be achieved by exploiting the difference in mean velocity, which is determined by the particle size. The optimal rate of phase change that maximizes the difference in mean velocity can be obtained from $\sqrt{|\rho^\ast|}=1$ at which the velocity ratio $\overline{v}/\overline{V}$ has the steepest gradient. By substituting \cref{eqn:ndVariables} into $\sqrt{|\rho^\ast|}=1$, the optimal rate of phase modulation for separation can be determined as
\begin{equation}\label{eqn:optphidot}
\dot \phi_{\rm{opt}} = \frac{2kF_0}{b}.
\end{equation}

The dimensionless period $\tau^\ast$ can be estimated by multiplying $\tau$ by $\dot\phi$, as the dimensionless time is scaled by $\dot\phi$ in \cref{eqn:ndVariables}. For particles larger than the target size, the period becomes infinite, indicative of no oscillatory movement [\Cref{fig:fig1_normalizedPeriod}]. The exact value of optimal rate of phase change cannot be calculated from \cref{eqn:optphidot} because of the difficulty in determining the acoustic pressure $p_0$. Instead, it can be estimated indirectly by monitoring oscillatory movements at decreasing $\dot \phi$.
The motion of target particle with respect to the rate of phase modulation can be predicted from the nondimensional analysis. When we let $a=R$ in \cref{eqn:ndVariables},
\begin{equation}
\sqrt{\left| {{\rho }^{*}} \right|}=\sqrt{\left| \frac{2k{{F}_{0}}}{b\dot{\phi }} \right|}. 				
\end{equation}
Primes in the equation were removed to indicate the motion of the target particle. As the rate of phase modulation $\dot{\phi}$ decreases from infinity, $\sqrt{\left| {{\rho }^{*}} \right|}$ increases from 0. When $\dot{\phi}$ is equal to the optimal rate, $\sqrt{\left| {{\rho }^{*}} \right|}=1$. It means that the period of oscillation becomes infinite and the oscillatory motion is diminished to zero as shown in \Cref{fig:fig1_normalizedPeriod}. Therefore, we can determine the optimal rate of phase change by observing reduction in oscillatory motion of target particle as the rate of phase change decreases from infinity.

\begin{figure}[!t]
	\centering
	\includegraphics[width=\columnwidth]{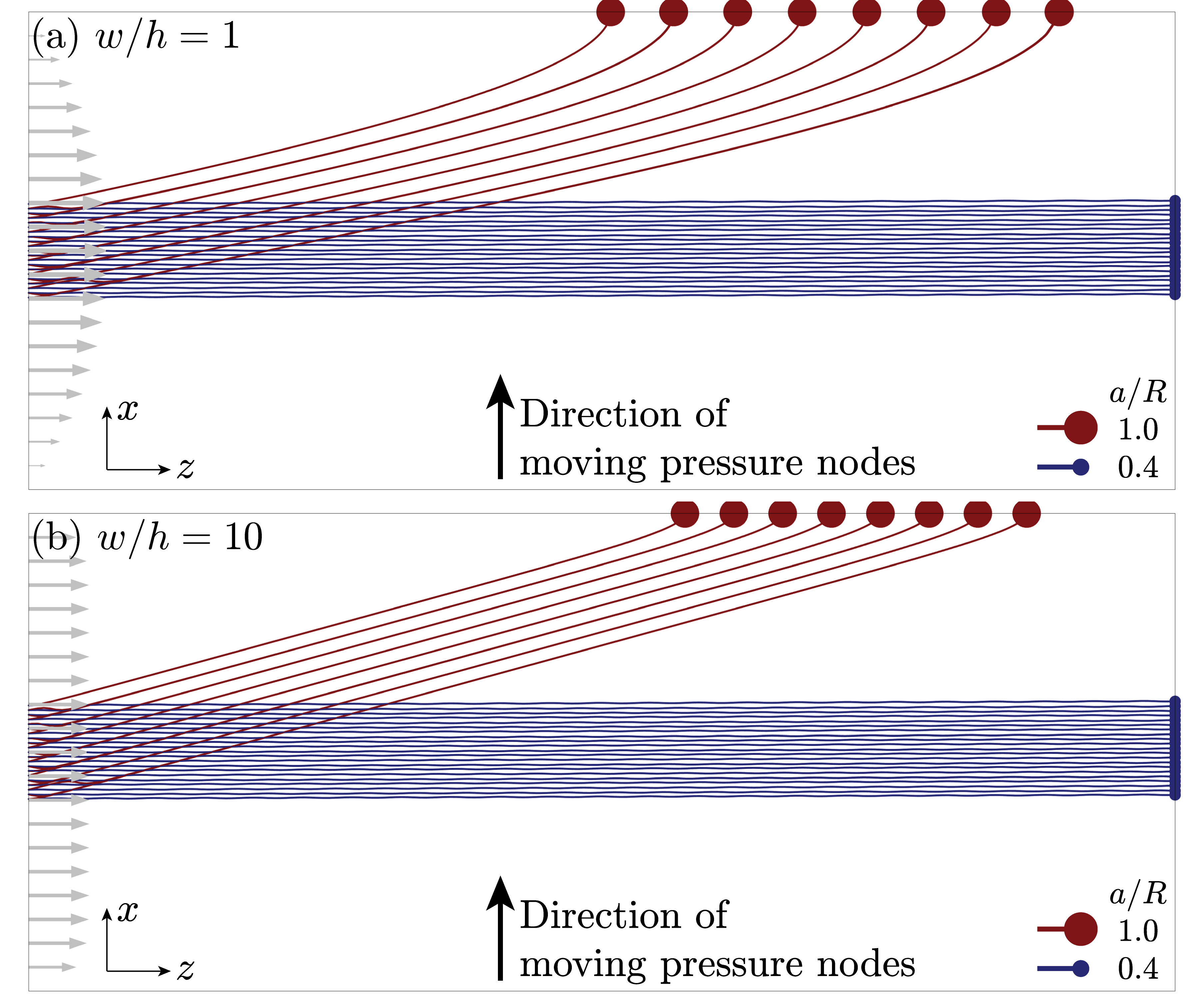}%
	\begin{subfigure}{0\columnwidth}%
		\phantomcaption%
		\label{fig:fig2_trajLow}%
	\end{subfigure}%
	\begin{subfigure}{0\columnwidth}%
		\phantomcaption%
		\label{fig:fig2_trajHigh}%
	\end{subfigure}%
	\caption{
		Trajectories of particles of radius $R$ (orange circles) and $0.4R$ (blue circles) subject to CPM-SSAW displacing pressure nodes in the positive X-direction.
		Regardless of the aspect ratio $\left(w/h\right)$ of a channel, target particles are successfully separated as pressure nodes are continuously translated by phase modulation in such wide channels.
		\subref{fig:fig2_trajLow} When a channel has a low aspect ratio, the flow profile becomes more curved, leading to a significant deflection in the trajectories.
		\subref{fig:fig2_trajHigh} The uniform flow profile formed in a channel with a high aspect ratio causes bead trajectories to become more straight.
	}
\end{figure}

Widening a microfluidic channel alters its aspect ratio, and accordingly the flow profile.
To predict the displacement of particles subjected to CPM-SSAW in a channel , we perform a finite element method using the CFD and Particle Tracing Modules in COMSOL Multiphysics\textsuperscript{\textregistered}.
We assume the Poisuille flow in a channel. Particles with target and non-target size are randomly distributed around the center region in the inlet.
Meshes for particle tracing are rectangular and dense enough to resolve the displacement of beads by acoustic radiation force on the length-scale of micrometers. The width of channel is much larger than the wavelength, and the width-to-height ratio varies to see how the dimension of channel affects particle separation using CPM-SSAW. 
The computational results show that regardless of the aspect ratios, CPM-SSAW is able to separate particles of the target size $R$ in a microfluidic channel with a width spanning multiple pressure nodes [Figs. \ref{fig:fig2_trajLow}, and \subref{fig:fig2_trajHigh}].
Trajectories of target particles are deflected to the direction of moving pressure nodes by CPM-SSAW, whereas those of non-target particles are rarely influenced following the streamlines of flow and remain dispersed.
We note that the trajectories of the target particles deflect more significantly in the vicinity of the channel when its width-to-height ratio is small. This is due to the parabolic flow profile in which the longitudinal velocity decreases more rapidly near the wall.

\begin{figure}[!t]
	\centering
	\includegraphics[width=\columnwidth]{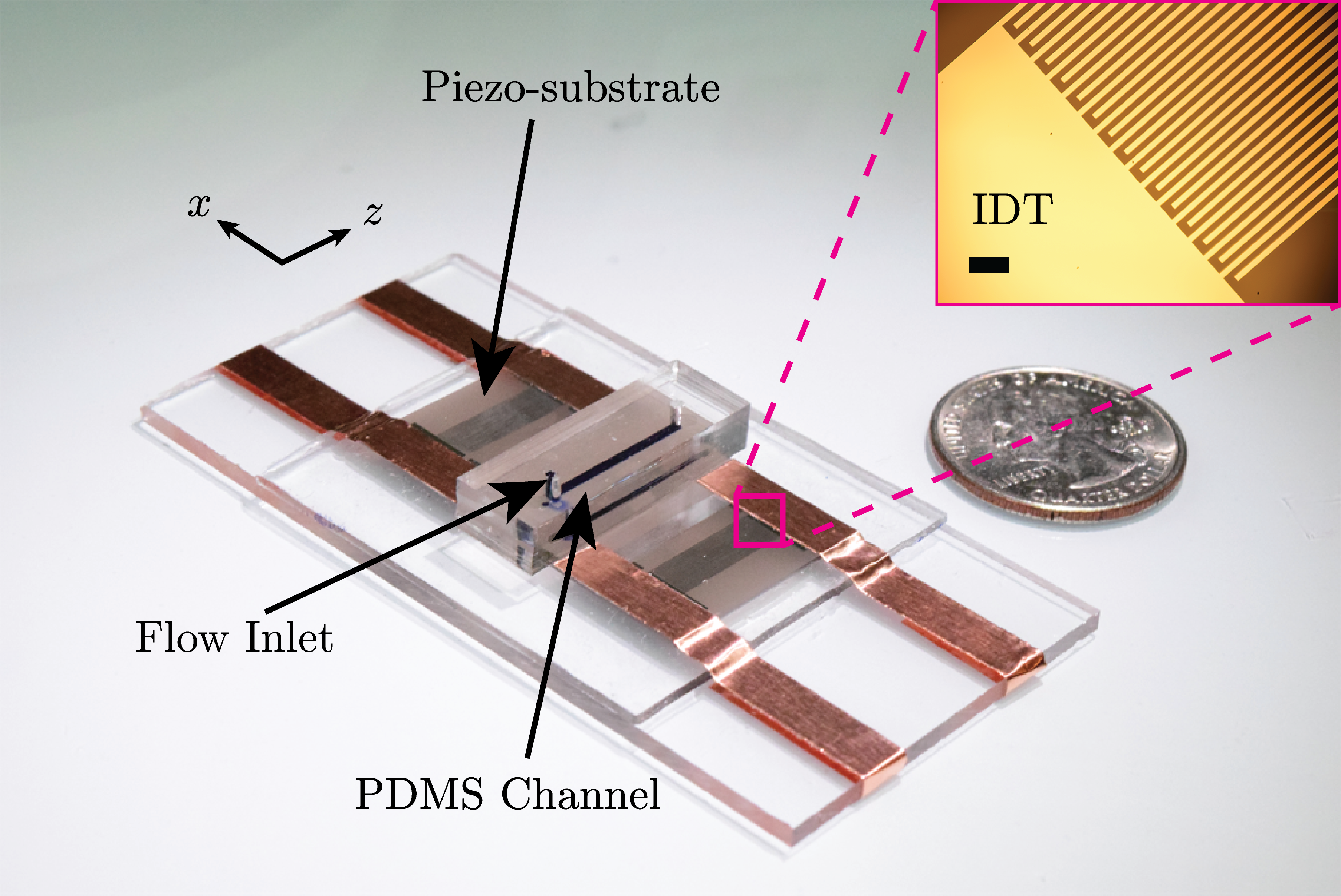}
	\caption{		
		Photo of the chip used in experiments for CPM-SSAW, showing the flow inlet, piezo-substrate, PDMS channel and IDTs. The \emph{inset} shows the zoomed image of IDTs (scale bar: \SI{500}{\micro\meter}).\label{fig:fig3_chipPhoto}
	}
\end{figure}
\begin{figure*}[!t]
	\centering
	\includegraphics[width=2\columnwidth]{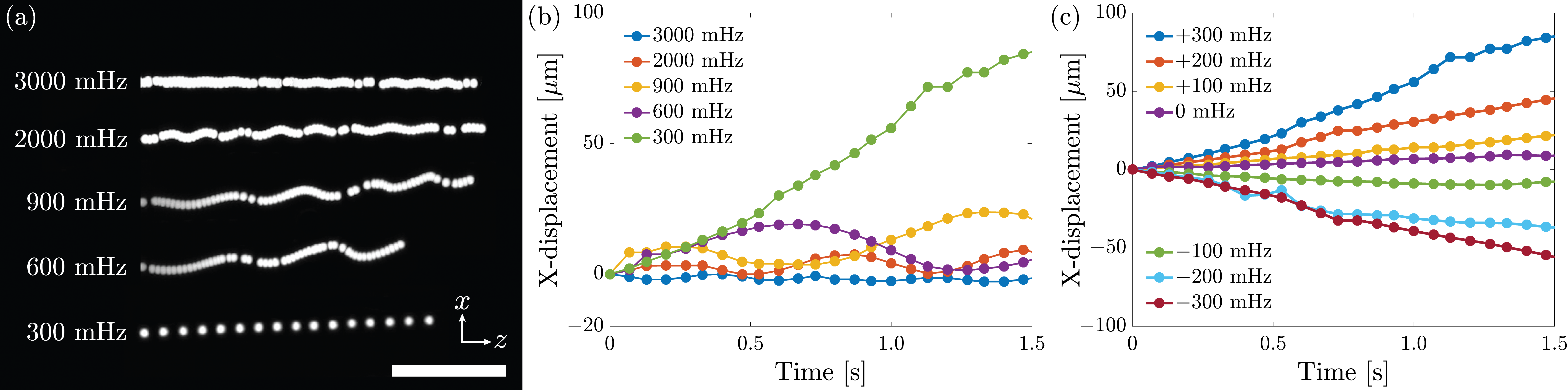}
	\begin{subfigure}{0\columnwidth}
		\phantomcaption
		\label{sFig:sFig_osc-exp}
	\end{subfigure}  
	\begin{subfigure}{0\columnwidth}
		\phantomcaption
		\label{sFig:sFig_osc-track}
	\end{subfigure}  
	\begin{subfigure}{0\columnwidth}
		\phantomcaption
		\label{sFig:sFig_trans-track}
	\end{subfigure}
	\caption{
		\textbf{Responses of the target particle of \SI{15}{\micro\meter} in diameter to various frequency shifts}
		\subref{sFig:sFig_osc-exp} A stack of fluorescent particle images showed that the target particle exhibits a linear or oscillatory movement depending on the magnitude of frequency shift.
		\subref{sFig:sFig_osc-track}, \subref{sFig:sFig_trans-track} X-directional displacement of the target particle when the frequency shift is higher and smaller than the optimal one of \SI{300}{\milli\hertz}.
		}
\end{figure*}
\begin{figure*}[!t]
	\centering
	\includegraphics[width=2\columnwidth]{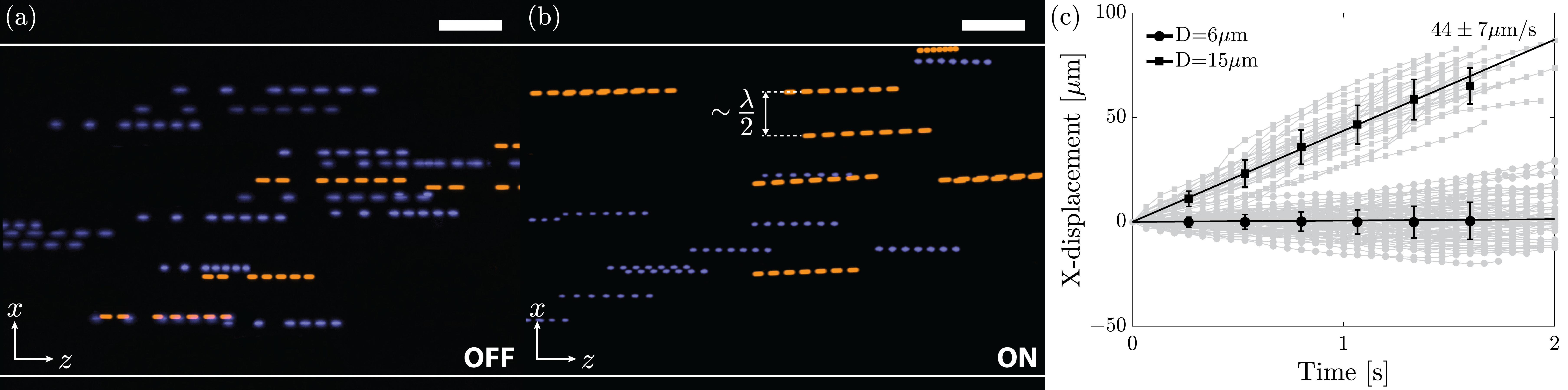}
	\begin{subfigure}{0\columnwidth}
		\phantomcaption
		\label{fig:fig3_stackingOff}
	\end{subfigure}
	\begin{subfigure}{0\columnwidth}
		\phantomcaption
		\label{fig:fig3_stackingOn}
	\end{subfigure}
	\begin{subfigure}{0\columnwidth}
		\phantomcaption
		\label{fig:fig3_tracking}
	\end{subfigure}
	\caption{
		\textbf{Separation of the target particles from non-target ones using CPM-SSAW.}
		\subref{fig:fig3_stackingOff} A stack of seven fluorescent images  of particles of \SI{6}{\micro\meter} (blue) and \SI{15}{\micro\meter} (orange) in diameter (scale bar: \SI{200}{\micro\meter}). The particles are infused to the right in a channel defined by white lines. Both particles travel along the flow, maintaining straight paths.
		\subref{fig:fig3_stackingOn} When the pressure nodes are translated upward by the linear phase modulation, the trajectories of \SI{15}{\micro\meter} particles are inclined toward the upper wall of the channel. In contrast, the paths of the \SI{6}{\micro\meter} particles are rarely affected by CPM-SSAM, and they remain relatively straight along the flow direction. The spaces between trajectories of \SI{15}{\micro\meter} particles are equal to a half-wavelength. 
		\subref{fig:fig3_tracking} X-displacement of particles of \SI{6}{\micro\meter} ($n=102$) and \SI{15}{\micro\meter} ($n=30$) in diameter. The averaged velocity for \SI{15}{\micro\meter} particles is $44\pm7$\si{\micro\meter\per\second}, which is fairly close to the speed of the moving pressure nodes, $\lambda/2 \times \dot\phi/2\pi = $\SI{42.8}{\micro\meter\per\second}. The error bar represents standard deviation of mean.
	}
\end{figure*}

The developed technique of particle separation using CPM-SSAW was also demonstrated experimentally. A PDMS micro-channel (w$\times$h: $\SI{1050}{\micro\meter}\times\SI{80}{\micro\meter}$) was molded using the soft lithography technique \cite{xia98,ha15}. IDTs (Cu/Al) were deposited on the piezoelectric substrate of \SI{128}{\degree} rotated Y-cut X-propagating lithium niobate. Each IDT had 23 finger pairs to produce CPM-SSAW with a wavelength of \SI{285}{\micro\meter} at a frequency of \SI{14}{\mega\hertz}. The PDMS micro-channel was bonded to the piezoelectric substrate through O\textsubscript{2} plasma treatment [\Cref{fig:fig3_chipPhoto}].
To control the phase of each wave, a sinusoidal voltage was generated in individual channels of a function generator (Keysight, HP33522), amplified by separate power amplifiers (Mini Circuits, LZY-22Y+), and supplied to each IDT independently.
Bead-bead and cell-bead mixtures were infused into a microchannel by a syringe pump (Chemyx, Fusion 200) at a flow rate of \SI{180}{\micro\liter\per\hour}.
Particles and cells were imaged by a fluorescence microscope (Nikon, Eclipse Ni-U) at the speed of 15 frames/s.

The bead-bead mixture was prepared by mixing fluorescent polystyrene microparticles of \SI{6}{\micro\meter} (blue, Polyscience BioMag\textsuperscript{\textregistered}) and \SI{15}{\micro\meter} (orange, Molecular Probe FluoSpheres\textsuperscript{\textregistered}) in diameter.
For preparing the cell-bead mixture, Human Keratinocytes (HaCaT) were mixed with \SI{2}{\micro\meter}-diameter polystyrene microparticles.
Cells were cultured in Dulbecco's modified eagle's medium (DMEM; Welgene) supplemented with 10\% fetal bovine serum (FBS; Invitrogen), 1\% penicillin/streptomycin solution (Pen/Strep; Invitrogen) and 1\% MEM non-essential amino acids (MEM NEAA; Invitrogen). They were subcultured every two days in 75T flasks and incubated for \SI{24}{\hour} at \SI{37}{\degreeCelsius}, 5\% CO\textsubscript{2} in culture media.
Cells were detached from the substrate by 0.05\% Trypsin-EDTA (Gibco) treatment, washed twice with 1x PBS, and fixed in 4\% paraformaldehyde (PFA; Electron Microscopy Science) for \SI{30}{\minute}.
For experiments, the fixed cells were resuspended with 1x PBS, and mixed with a solution of \SI{2}{\micro\meter}-diameter polystyrene microparticles. 

A linear continuous phase modulation can be achieved by shifting a frequency in a small amount, because $y=y_0 \cos(kx-\omega t-\dot \phi t) = y_0 \cos(kx-(\omega+\dot\phi)t)$ \cite{tran12}.
Therefore, we altered the frequency of one channel of the function generator in our experiments to move pressure nodes along the direction perpendicular to the flow.
Positions of particles were tracked by acquiring images continuously during experiments. A mean velocity of particles was calculated by fitting the displacement to a line and taking an average of slopes of the fitted lines \cite{schindelin12}.

The optimal value of frequency shift, or the optimal rate of phase change, was determined by monitoring oscillatory motion of target particles of \SI{15}{\micro\meter} in diameter.
It was determined to be \SI{300}{\milli\hertz} at which the target particles stopped exhibiting any oscillatory movement while decreasing the frequency shift from \SI{3000}{\milli\hertz}.
We found the target particle exhibited oscillatory or linear displacement depending on the magnitude of frequency shift [\Cref{sFig:sFig_osc-exp}].
To see how the force by CPM-SSAW contributes in the bead displacements more clearly, X-directional components of the bead displacement was analyzed.
When the frequency shift is higher than the optimal value of \SI{300}{\milli\hertz}, particles failed to be displaced with the moving pressure nodes and instead showed oscillatory movements [\Cref{sFig:sFig_osc-track}].
At the frequency shift smaller than the optimal value, particles are displaced without any oscillatory movement [\Cref{sFig:sFig_trans-track}].
The speed of bead displacement is proportional to the magnitude of frequency shift.

We applied the CPM-SSAW with the optimized frequency shift to separate particles with the target size from others in a bead mixture. 
When CPM-SSAW was off, both target and non-target particles traveled straight along the streamline of flow [\Cref{fig:fig3_stackingOff}].
When CPM-SSAW was applied, we observed that only particles with the target size of \SI{15}{\micro\meter} in diameter were displaced toward the wall of the channel without any oscillatory motion [\Cref{fig:fig3_stackingOn}] (Supplementary Video).
In contrast, the trajectories of \SI{6}{\micro\meter}-diameter particles remained relatively straight indicative of being rarely displaced by CPM-SSAW. We noted that the number of \SI{15}{\micro\meter} particles sticking to the wall continued to increase during experiment as a result of the selective bead separation. The \SI{15}{\micro\meter}-diameter particles were spaced equally by a half-wavelength, \SI{148}{\micro\meter}. The speed of the targeted beads was estimated to be $44\pm7$\si{\micro\meter\per\second}, which agreed well with that of the moving pressure nodes, \SI{42.8}{\micro\meter\per\second} [\Cref{fig:fig3_tracking}].
\begin{figure}[!t]
	\centering
	\includegraphics[width=\columnwidth]{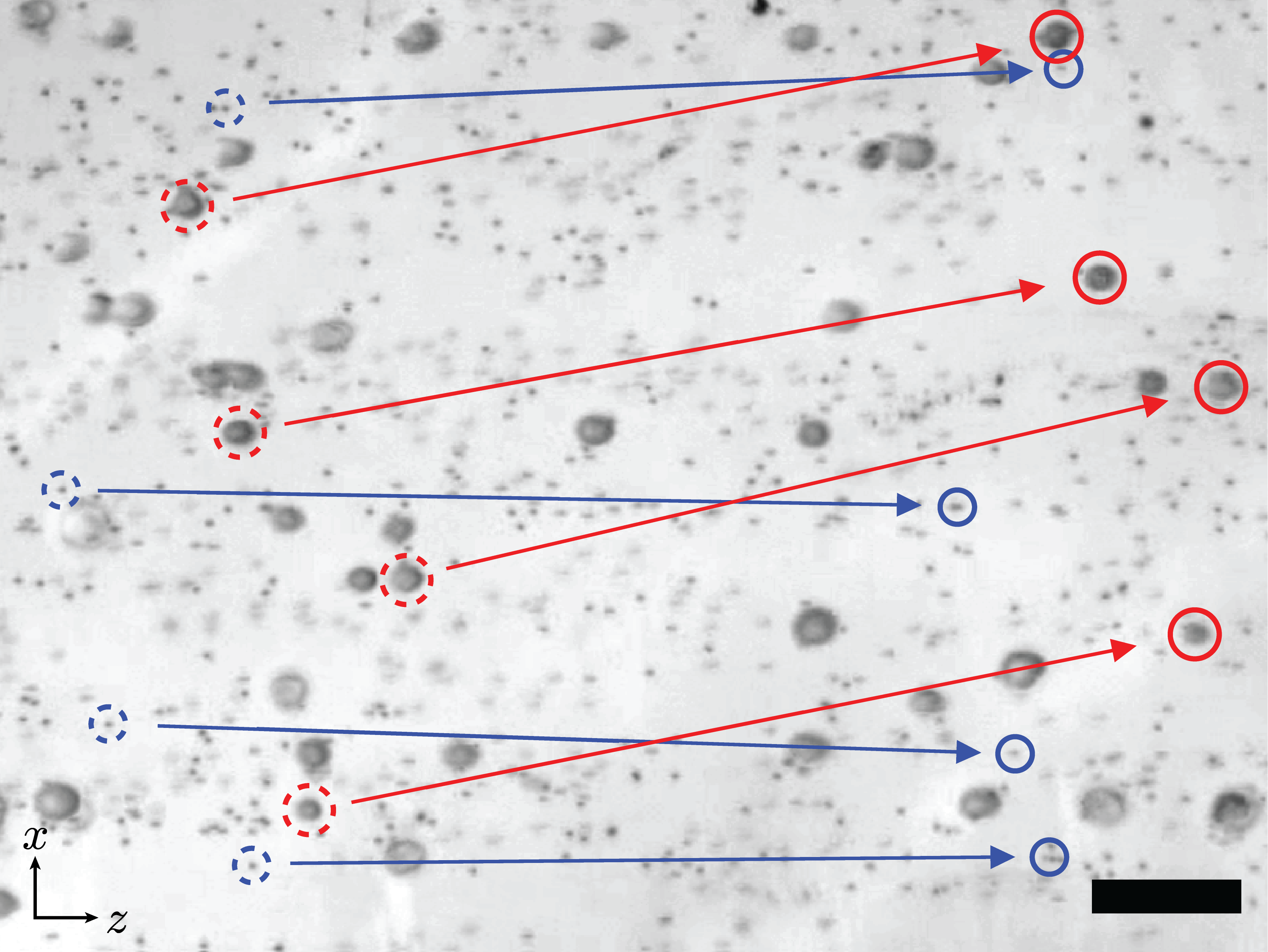}
	\caption{		
		\textbf{Separation of cells from particles using CPM-SSAW.} Superposition of images of HaCaT cells (Red) and \SI{2}{\micro\meter}-diameter beads (Blue) subject to CPM-SSAW at $t=0$ (Dash-outlined circle) and $t=\SI{5}{\second}$ (Solid-outlined circle) (Scale bar: \SI{100}{\micro\meter}).
		\label{fig:cellSeparation}
	}
\end{figure}
The separation selectivity was manifested in the experiments for a cell-bead mixture.
The width of channel was \SI{1050}{\micro\meter}, which is large enough to contain multiple pressure nodes.
As demonstrated in \Cref{fig:cellSeparation}, our CPM-SSAW technique was able to displace HaCaT cells along the direction of moving pressure nodes while it rarely affected the trajectories of \SI{2}{\micro\meter}-diameter particles.
This result suggests that our technique can be advantageous in developing a massive separation method for cells without labels.  

By using CPM-SSAW, we develop a particle separation method for channels that are wider than a wavelength. When SSAW are translated by phase modulation, forces applied to the particles are determined by their size and contrast factors. Therefore, continuous phase modulation can cause particles of a specific dimension to be displaced along the direction of moving pressure nodes, while others remain still or move slowly with an oscillatory motion. Through experiments and computational simulations, we demonstrated that CPM-SSAW is able to separate micro-sized substances such as particles and cells of a specific size regardless of their locations.
As the separation technique using CPM-SSAW can be applied to multiple pressure nodes, it will be beneficial for developing a high-throughput separation method with high selectivity.

\section{Acknowledgements}
This work was supported by the convergence technology development program for bionic arm through the National Research Foundation of Korea funded by the Ministry of Science, ICT \& Future Planning (NRF-2014M3C1B2048449) and the Yonsei University Future-leading Research Initiative of 2014 (RMS 2015-22-0168).

\balance

\setlength\bibsep{0.2\itemsep}
\bibliography{CPMSSAW}
\end{document}